\documentstyle[preprint,revtex,eqsecnum]{aps}
\begin{document}
\draft
\preprint{UAHEP-9214}
\preprint{August 1992}
\begin{title}
Statistical Mechanics of Extended Black Objects
\end{title}
\author{B. Harms and Y. Leblanc}
\begin{instit}
Department of Physics and Astronomy, The University of Alabama\\
Box 870324, Tuscaloosa, AL 35487-0324
\end{instit}
\begin{abstract}
We extend the considerations of a previous paper on black hole
statistical mechanics to the case of black extended objects such
as black strings and black membranes in 10-dimensional
space-time. We obtain a general expression for the Euclidean
action of quantum black p-branes and derive their corresponding
degeneracy of states. The statistical mechanics of a gas of black
p-branes is then analyzed in the microcanonical ensemble. As in
the case of black holes, the equilibrium state is not thermal and
the stable configuration is the one for which a single black
object carries most of the energy. Again, neutral black p-branes
obey the bootstrap condition and it is then possible to argue
that their scattering amplitudes satisfy crossing symmetry.
Finally, arguments identifying quantum black p-branes with
ordinary quantum branes of different dimensionality are
presented.
\end{abstract}

\narrowtext
\section{INTRODUCTION}
\label{sec:intro}

A complete picture of black hole physics (including the effects
of quantum mechanics) is still missing at present. This problem
is very much at the core of a larger problem, that of a
consistent description of a quantum theory of gravity. However,
as is well known, to this day, the only consistent theory of
quantum gravity (as well as that of a comprehensive unification
of all forces) is string theory. Therefore, it may not be
surprising to encounter string physics in black hole physics.

t'Hooft, in a series of inspiring articles \cite{tH90}, has led
the way for the search of a proper particle Hilbert space in the
presence of   black holes. The effect of a particle falling into
or escaping from the horizon of a Schwarzschild black hole is
actually to shift the horizon and thereby produce a so-called
gravitational shock wave \cite{Dt85}. The corresponding
scattering matrix elements (built on the Hilbert space of the
particles' momenta) have been shown to be formally identical with
those of a string theory with imaginary string tension
\cite{tH90,Ve91}.

In a recent paper \cite{Co92}, the present authors in
collaboration with Paul Cox generalized the above considerations
to the case of a dilaton black hole and found that the
requirement of real string tension (unitarity) could be achieved
for a specific value of the dilaton parameter in the naked
singularity domain $r_->r_+$. This result seems to indicate that
quantum black holes can become (or decay into) quantum strings
beyond the extreme point $r_-=r_+$, a fact which may find its
experimental verification with the discovery of extragalactic
cosmic gamma ray bursts \cite{Di88}. However this also indicates
that black holes are not quite strings, an assertion supported by
the asymptotic behavior of their density of states which grows
faster than that of strings as mass increases.

In a more traditional approach, following semiclassical
treatments in which the above gravitational back reaction effects
were neglected, the area of a black hole event horizon has been
interpreted in terms of thermodynamical entropy and the black
hole mass related to a canonical temperature called the
Bekenstein-Hawking temperature \cite{Be73,Ha75}. This picture
however, as was pointed out in previous articles
\cite{CPW91,Pr91,SC91,Ho91,BL92}, leads to difficulties with both
thermodynamics and quantum mechanics. This is especially so in
the thermally interpreted process of black hole evaporation as
pure states are converted into mixed states. Attempts to resolve
this problem by taking into account the effects of quantum hair
have been made \cite{CPW91,Pr91,SC91,Ho91}.

There exists however an alternative interpretation of the
semiclassical (WKB) calculation which does not violate the laws
of quantum mechanics \cite{BL92}. Namely, the saddle point
approximation to the path integral with Euclidean action can be
regarded as a tunneling probability per unit volume of a particle
escaping the event horizon. This is basically a quantum
mechanical barrier penetration problem and so the tunneling
probability is an effective measure of the ratio of a single
particle state escaping the black hole to the number of available
states ${\rho}_{B.H.}$ inside (and including) the horizon. We
then arrive at the following approximate semiclassical formula
for the  black hole degeneracy of states at mass level m,
\begin{eqnarray}
{\rho}_{B.H.}(m)\;\simeq\;c\,\exp{{S_E(m)}\over\hbar}\;,
\end{eqnarray}
in which the constant c represents general quantum field
theoretical corrections and $S_E$ is the Euclidean action
(so-called Bekenstein-Hawking entropy) of the classical solutions
(instantons) of the Euclidean equations of motion
\cite{GPY81,GPY82}. These instantons are actually periodic
instantons as the condition of the vanishing of the conical
singularity of the Euclidean space-time in the black hole
background requires Euclidean time $\tau$ to be a compact space
($S^1$) with period $\beta_H$ (the inverse Hawking temperature).
The integral over Euclidean time in Eq.(1.1) is to be evaluated
over a single period and, making use of the relation between
$\beta_H$ and the black hole mass m, the density  $\rho_{B.H.}$
becomes solely a function of mass (and possibly electric charge,
angular momentum, etc.). In the case of a 4-dimensional
Schwarzschild black hole, one obtains ($\hbar=c=G=1$),
\begin{eqnarray}
{\rho}_{Schw}(m)\;\sim\;c\,e^{4\pi{m^2}}\;,
\end{eqnarray}
a result to be compared with the softer behavior for strings,
\begin{eqnarray}
{\rho}_{string}(m)\;\sim\;c\,{m^a}\,e^{bm}\;.
\end{eqnarray}
However, like strings, Schwarzschild black holes have been shown
\cite{BL92} to obey the statistical bootstrap condition
\cite{Hg70,Fr71,Ca72}. Furthermore, arguments were presented
\cite{BL92} which made it very plausible for the scattering
amplitudes to obey duality symmetry. That these properties were
realized for such a case originated from the fact that extreme
Schwarzschild black holes are massless. Quantum black holes may
then belong to some class of conformal theories, perhaps
p-branes. Electric or magnetic charge (hair) tends to destroy
these properties \cite{BL92}.

In this work, we generalize these analyses to cases of quantum
black extended objects such as black strings and black membranes
in 10-dimensional space-time. Our considerations are based on
recent findings by Horowitz and Strominger \cite{HS91} as well as
Gibbons and Maeda \cite{GM88}.

We first obtain an explicit general expression for the degeneracy
of states of black (10-D)-branes ($4\leq D\leq 10$) and then
proceed to analyze the statistical mechanics of a gas of such
objects making use of the microcanonical ensemble. As in the case
of black holes, the canonical partition function of black
(10-D)-branes diverges for all temperatures. Our results closely
resemble those for dilaton black holes \cite{BL92}.

Although the canonical partition function is formally divergent,
a fact due to the negative microcanonical specific heat, it may
be possible to extract information with respect to the nucleation
rate (decay rate per unit volume) of a gas of black objects by
evaluating this same partition function in the convergence domain
of specific parameters appearing in the expression for the
density of states. Once the integral is performed, these
parameters are then analytically continued back to their original
values, an operation which often generates imaginary terms. The
imaginary part of the corresponding free energy is then simply
related, as has been shown by Langer in condensed matter systems
\cite{La6769}, to the decay rate of the metastable black objects
gaseous phase.

\section{BLACK (10-D)-BRANES}
\label{sec:black}

In this section, we proceed to derive a general expression for
the degeneracy of states of a quantum black (10-D)-brane
($4\leq D\leq 10$) making use of the semiclassical (WKB)
approximated expression for the tunneling probability per unit
volume. In complete parallel to the black hole case (cf.
Eq.(1.1)), we have,
\begin{eqnarray}
{\rho}_{B.B.}(m)\;\simeq\;c\,\exp{{S_E(m)}\over\hbar}\;.
\end{eqnarray}

Following Horowitz and Strominger \cite{HS91}, we wish to
consider the following 10-dimensional action,
\begin{eqnarray}
S\;=\;{1\over{16\pi}}\,\int{{d^{10}}x}\,\sqrt{-g}\,\biggl[\,e^{-
2\Phi}\,\bigl[\,R\;+\;4\,(\partial\Phi)^2\,\bigr]\;-
\;{{2\,e^{2\alpha\Phi}}\over{(D\,-\,2)!}}\,F^2\,\biggr]\;,
\end{eqnarray}
in which the field $F_{{\mu_1}...{\mu_{D-2}}}$ is a (D-2)-form
satisfying $dF\,=\,0$ and from which a magnetic charge
$Q\,\propto\,\int F$ is carried by objects spatially extended in
(10-D)  dimensions. In the above action, the field $\Phi$
represents the dilaton and R is the scalar curvature of the 10-
dimensional space-time. Actions similar to this one are found in
string theories. The search for charged black (10-D)-brane
solutions extremizing the action (2.2) has been considerably
simplified by reducing the problem to finding dilaton black hole
solutions of an effective D-dimensional action \cite{HS91}. Such
a solution has been given by Gibbons and Maeda \cite{GM88}.

Through field redefinitions, Horowitz and Strominger arrived at
the following action,
\begin{eqnarray}
S\;=\;{{L^{10-D}}\over{16\pi}}\,\int{{d^D}x}\,\sqrt{-
g}\,\biggl[\,{\hat R}\;-\;{1\over2}\,{(\nabla\phi)}^2\;-
\;{{2\,e^{-a\phi}}\over{(D\,-\,2)!}}\,F^2\,\biggr]\;,
\end{eqnarray}
in which $L^{10-D}$ is the volume of (10-D) space, ${\hat
g}_{\mu\nu}$ is the induced metric of D-dimensional space-time
with Riemann curvature ${\hat R}_{\mu\nu}$, $\phi$ is the
rescaled dilaton field and the dilaton parameter $a$ is given by,
\begin{eqnarray}
a\;\equiv\;\sqrt{4{\alpha^2}\,+\,2\alpha(7-D)\,+\,2{(D-1)\over(D-
2)}}\;.
\end{eqnarray}

The equations of motion derived by extremizing the action (2.3)
are obtained as follows \cite{HS91},
\begin{eqnarray}
{\nabla^{\mu_1}}[\,e^{-a\phi}{F_{{\mu_1}\cdots{\mu_{D-
2}}}}\,]\;=\;0\;,
\end{eqnarray}
\begin{eqnarray}
{\nabla^2}\phi\;=\;-{{2\alpha}\over{(D-2)!}}\,e^{-a\phi}F^2\;,
\end{eqnarray}
and,
\begin{eqnarray}
{\hat R}_{\mu\nu}\,-
\,{1\over2}{\nabla_{\mu}}\phi{\nabla_{\nu}}\phi\;&=&\;{2\over{(D-
3)!}}\,e^{-a\phi}{F_{{\mu}{\lambda_1}\cdots{\lambda_{D-
3}}}}{F_{\nu}^{{\lambda_1}\cdots{\lambda_{D-3}}}}\nonumber\\
&&-2{\hat g}_{\mu\nu}{(D-3)\over{(D-2)(D-2)!}}\,e^{-a\phi}F^2\;.
\end{eqnarray}

Spherically symmetric solutions describing dilaton black holes in
D dimension have been obtained \cite{HS91,GM88},
\begin{eqnarray}
d{\hat s}^2\;=\;-e^{2\Phi(\hat r)}d{t^2}\,+\,e^{2\Lambda(\hat
r)}d{{\hat r}^2}\,+\,{{R^2}(\hat r)}d{\Omega_{D-2}^2}\;,
\end{eqnarray}
\begin{eqnarray}
{F_{{\mu_1}\cdots{\mu_{D-
2}}}}\;&=&\;Q\,\epsilon_{{\mu_1}\cdots{\mu_{D-2}}}\;;\nonumber\\
&& \\
{{F^2}\over{(D-2)!}}\;&=&\;{{Q^2}\over{{R^{2(D-2)}}(\hat
r)}}\;,\nonumber
\end{eqnarray}
and,
\begin{eqnarray}
e^{-a\phi(\hat r)}\;=\;{\biggl[1\,-\,{{\biggl({{r_-
}\over{r}}\biggr)}^{D-3}}\biggr]}^{\gamma(D-3)}\;,
\end{eqnarray}
where,
\begin{eqnarray}
e^{2\Phi(\hat r)}\,=\,e^{-2\Lambda(\hat r)}\,=\,\biggl[1\,-
\,{{\biggl({{r_+}\over{r}}\biggr)}^{D-3}}\biggr]{\biggl[1\,-
\,{{\biggl({{r_-}\over{r}}\biggr)}^{D-3}}\biggr]}^{1-\gamma(D-
3)}\;,
\end{eqnarray}
\begin{eqnarray}
{{R^2}(\hat r)}\;=\;{r^2}\,{\biggl[1\,-\,{{\biggl({{r_-
}\over{r}}\biggr)}^{D-3}}\biggr]}^{\gamma}\;,
\end{eqnarray}
\begin{eqnarray}
\gamma\;=\;{{2{a^2}(D-2)}\over{(D-3)[\,2(D-3)\,+\,{a^2}(D-
2)\,]}}\;,
\end{eqnarray}
and where ${\hat r}\,=\,{\hat r}(r)$ can be obtained from the
relation \cite{HS91},
\begin{eqnarray}
{r^{D-4}}dr\;=\;{R^{D-4}}d{\hat r}\;.
\end{eqnarray}

The above solutions are parametrized by two horizons situated at
$r_+$ and $r_-$, parameters which can be expressed in terms of
the charge Q and mass M of the D-dimensional dilaton black hole
as follows,
\begin{eqnarray}
{Q^2}\;=\;{{\gamma{(D-3)^3}{({r_+}{r_-})}^{D-3}}\over{2{a^2}}}\;,
\end{eqnarray}
and,
\begin{eqnarray}
M\;=\;{{(D-2){\pi^{{D-3}\over2}}}\over{8\Gamma({{D-
1}\over2})}}\,\bigl[\,{r_+^{D-3}}\,+\,[1-\gamma(D-3)]{r_-^{D-
3}}\,\bigr]\;.
\end{eqnarray}

These relations can be inverted to yield,
\begin{eqnarray}
{r_+^{D-3}}\;=\;{{4\Gamma({{D-1}\over2})M}\over{(D-2){\pi^{{D-
3}\over2}}}}\,\biggl[1\,+\,\sqrt{1\,-\,{{{a^2}{{(D-
2)}^2}{Q^2}{\pi^{D-3}}[1-\gamma(D-3)]}\over{8\gamma{{(D-
3)}^3}{\Gamma^2}({{D-1}\over2}){M^2}}}}\biggr]\;,
\end{eqnarray}
and,
\begin{eqnarray}
{r_-^{D-3}}\,=\,{{{a^2}{Q^2}(D-2){\pi^{{D-
3}\over2}}}\over{2\gamma{(D-3)^3}\Gamma({{D-
1}\over2})M}}{{\biggl[1\,+\,\sqrt{1\,-\,{{{a^2}{{(D-
2)}^2}{Q^2}{\pi^{D-3}}[1-\gamma(D-3)]}\over{8\gamma{{(D-
3)}^3}{\Gamma^2}({{D-1}\over2}){M^2}}}}\biggr]}^{-1}}\;.
\end{eqnarray}

Now in order to compute the semiclassical approximation of the
path integral, the above solutions must be analytically continued
to Euclidean time $\tau$. In the Euclidean space-time, they
become instanton solutions. However, in this Euclidean
formulation, a surface term ($S_{bd}$) [21] must be added to the
analytical continuation of the action (2.2), which we denote by
$S_o$. Therefore the full Euclidean action is given by,
\begin{eqnarray}
{S_E}\;=\;{S_o}\,+\,{S_{bd}}\;.
\end{eqnarray}

In addition, requiring the absence of the conical singularity in
the Euclidean space-time yields the so-called inverse
Bekenstein-Hawking temperature \cite{Be73,Ha75,SC91},
\begin{eqnarray}
{\beta_H}\;=\;{{2\pi}\over{{\bigl[(\,{\partial_{\hat
r}}{e^{\Phi({\hat r})}}\,)\,{e^{-\Lambda({\hat
r})}}\bigr]}_{{\hat r}={\hat r}({r_+})}}}\;.
\end{eqnarray}
We find,
\begin{eqnarray}
{\beta_H}\;=\;{{4\pi{r_+}}\over{D-3}}\,{\biggl[1\,-
\,{{\biggl({{r_-}\over{r_+}}\biggr)}^{D-3}}\biggr]}^{{{\gamma(D-
2)}\over2}-1}\;.
\end{eqnarray}
This result can be re-expressed in terms of M and Q with the aid
of Eqs. (2.17) and (2.18). We get,
\begin{eqnarray}
{\beta_H}\;&=&\;{{4\pi}\over{D-3}}{{\biggl[{{4\Gamma({{D-
1}\over2})}\over{(D-2){\pi^{{D-3}\over2}}}}\biggr]}^{1\over{D-
3}}}\,{M^{1\over{D-3}}}\,{{\biggl(1\,+\,\sqrt{1\,-\,\lambda[1-
\gamma(D-3)]}\biggr)}^{1\over{D-3}}}\nonumber\\
&&\times\,{{\biggl[1\,-\,\lambda{{\biggl(1\,+\,\sqrt{1\,-
\,\lambda[1-\gamma(D-3)]}\biggr)}^{-2}}\biggr]}^{{{\gamma(D-
2)}\over2}-1}}\;;\nonumber\\
&&\\
\lambda\;&\equiv&\;{{{a^2}{Q^2}{(D-2)^2}{\pi^{D-
3}}}\over{8\gamma{(D-3)^3}{\Gamma^2}({{D-
1}\over2}){M^2}}}\;.\nonumber
\end{eqnarray}

The calculation of the contribution $S_o$ to the Euclidean action
is straightforward. Contracting Einstein's equation (Eq. (2.7))
with the metric tensor ${{\hat g}_{\mu\nu}}$ (after rotation to
Euclidean time), inserting the result into the action and making
use of the Eqs. (2.9)-(2.14),we arrive at the following
expression,
\begin{eqnarray}
{S_o}\;=\;-\,{{{L^{10-D}}{\pi^{{D-
3}\over2}}{Q^2}{\beta_H}}\over{4\Gamma({{D-
1}\over2})}}\biggl[1\,+\,{{D-4}\over{D-
2}}\biggr]\,\int_{r_+}^{\infty}dr\,{r^{-(D-2)}}\;,
\end{eqnarray}
in which the (D-4)-term originates from the non-vanishing trace
of the gauge field energy-momentum tensor for $D>4$. The above
action has been evaluated for a single period in the compact
Euclidean time. This is the contribution from the periodic
instantons. Carrying out the integral finally yields,
\begin{eqnarray}
{S_o}\;=\;-\,{{{L^{10-D}}{\pi^{{D-
3}\over2}}{Q^2}{\beta_H}}\over{2(D-2)\Gamma({{D-
1}\over2}){{r_+}^{D-3}}}}\;.
\end{eqnarray}

Now in order to evaluate the surface boundary action, we follow
the analyses of Ref.~\cite{SC91}. We have,
\begin{eqnarray}
{S^{\rm (boundary)}}\;=\;-\,{{L^{10-D}}\over{8\pi}}(D-
3)\,{[\,e^{-\Lambda({\hat r})}{\partial_{\hat
r}}{(Volume\,of\,boundary)}\,]}_{{\hat r}\rightarrow\infty}\;,
\end{eqnarray}
where,
\begin{eqnarray}
{Volume\,of\,boundary}\;=\;{\beta_H}e^{\Phi({\hat
r})}\,{{2{\pi^{{D-1}\over2}}}\over{\Gamma({{D-
1}\over2})}}\,{{\hat r}^{D-2}}\;.
\end{eqnarray}

The above action however is divergent in the limit ${\hat
r}\rightarrow\infty$ and so a subtraction must be performed,
namely that of a flat space contribution. Again following the
considerations of Ref.~\cite{SC91}, the flat space contribution
is obtained as follows,
\begin{eqnarray}
{S_{flat}^{\rm (boundary)}}\;=\;-\,{{L^{10-D}}\over{4\pi}}(D-
3)\,{\beta_H}e^{\Phi({\hat r})}\,{{\pi^{{D-
1}\over2}}\over{\Gamma({{D-1}\over2})}}\,{{\biggl[{\partial_{\hat
r}}\,({{\hat r}^{D-2}})\biggr]}_{{\hat r}\rightarrow\infty}}\;.
\end{eqnarray}

Therefore,
\begin{eqnarray}
{S_{bd}}\;=\;{S^{\rm (boundary)}}\,-\,{S_{flat}^{\rm
(boundary)}}\;.
\end{eqnarray}

Explicit evaluation finally yields,
\begin{eqnarray}
{S_{bd}}\;=\;{{(D-3){L^{10-D}}{\pi^{{D-
3}\over2}}{\beta_H}}\over{8\Gamma({{D-
1}\over2})}}\,\bigl[{r_+^{D-3}}\,+\,[1-\gamma(D-3)]{r_-^{D-
3}}\bigr]\;.
\end{eqnarray}

Making use of Eqs. (2.19), (2.24) and (2.29), the full Euclidean
action for the periodic instanton solutions is finally expressed
as follows,
\begin{eqnarray}
{S_E}\;=\;{{{L^{10-D}}{\pi^{{D-1}\over2}}}\over{2\Gamma({{D-
1}\over2})}}\,{{R^{D-2}}({\hat r})}\;\equiv\;{{\cal A}\over4}\;,
\end{eqnarray}
where $\cal A$ is the horizon area of the black (10-D)-brane. In
terms of M and Q, we get,
\begin{eqnarray}
{S_E}({\cal M},Q)\;&=&\;{{{L^{10-D}}{\pi^{{D-
1}\over2}}}\over{2\Gamma({{D-1}\over2})}}{{\biggl[{{4\Gamma({{D-
1}\over2})}\over{(D-2){\pi^{{D-3}\over2}}}}\biggr]}^{{D-
2}\over{D-3}}}\,{M^{{D-2}\over{D-3}}}\,{{\biggl(1\,+\,\sqrt{1\,-
\,\lambda[1-\gamma(D-3)]}\biggr)}^{{D-2}\over{D-3}}}\nonumber\\
&&\times\,{{\biggl[1\,-\,\lambda{{\biggl(1\,+\,\sqrt{1\,-
\,\lambda[1-\gamma(D-3)]}\biggr)}^{-2}}\biggr]}^{{\gamma(D-
2)}\over2}}\;,
\end{eqnarray}
where $\lambda$ has been defined in Eq. (2.22). This
is the Bekenstein-Hawking entropy. It is easy to check that,
\begin{eqnarray}
{{d{S_E}}\over{d{\cal M}}}\;=\;{1\over{L^{10-
D}}}\,{{d{S_E}}\over{dM}}\;=\;{\beta_H}\;,
\end{eqnarray}
where ${\cal M}\,\equiv\,M{L^{10-D}}$ is the total mass of the
black (10-D)-brane. Actually, the overall proportionality
constant in the expression (2.16) for the mass density M was
chosen in such a way that the relation (2.32) held.

According to Horowitz and Strominger \cite{HS91}, dual black
branes with electric charges can be obtained from the following
substitution,
\begin{eqnarray}
\alpha\,\rightarrow\,-\alpha\;\;;\;\;D\,\rightarrow\,14-D\;.
\end{eqnarray}

We close this section by providing the expression for the mass
density of the extreme black (10-D)-branes. It is derived from
the condition ${r_+}\,=\,{r_-}$. We find,
\begin{eqnarray}
{M_o}\;=\;{{\sqrt{D-2}{\pi^{{D-3}\over2}}Q}\over{2\Gamma({{D-
1}\over2})\sqrt{{a^2}(D-2)\,+\,2(D-3)}}}\;.
\end{eqnarray}

\section{STATISTICAL MECHANICS}
\label{sec:stat}

For small charge, according to eqs. (2.1) and (2.31), black
(10-D)-branes are characterized by the following degeneracy of
states,
\begin{eqnarray}
{\rho_{B.B.}}({\cal
M},Q)\;\simeq\;c\,\exp{\Bigl(\,\sigma(D)\,{M^{{D-2}\over{D-
3}}}\,\bigl[1\,+\,{\cal O}({Q^2}/{M^2})\bigr]\,\Bigr)}\;,
\end{eqnarray}
in which we defined,
\begin{eqnarray}
\sigma(D)\;\equiv\;{{{L^{10-D}}{\pi^{{D-
1}\over2}}}\over{2\Gamma({{D-
1}\over2})}}\,{{\biggl[{{8\Gamma({{D-1}\over2})}\over{(D-
2){\pi^{{D-3}\over2}}}}\biggr]}^{{D-2}\over{D-3}}}\;,
\end{eqnarray}
and the constant c represents the quantum field theoretical
effects.

The microcanonical density of states $\Omega({\cal E},Q,V)$ of a
gas of identically charged black objects of charge Q, degeneracy
${\rho_{B.B.}}({\cal M},Q)$, total energy $\cal E$ and enclosed
in a 9-dimensional volume V is expressed as follows,
\begin{eqnarray}
\Omega({\cal E},Q,V)\;=\;\sum_{n=1}^{\infty}\,{\Omega_n}({\cal
E},Q,V)\;,
\end{eqnarray}
in which the contribution from n black objects is given as,
\begin{eqnarray}
{\Omega_n}({\cal
E},Q,V)\;=\;{{\biggl[{V\over{{(2\pi)}^9}}\biggr]}^n}
&{1\over{\Gamma(n+1)}}\,\prod\limits_{i=1}^{n}\biggl[\,\int\limits_{{\cal
M}_o}^
M}_i},Q)\,\int\limits_{-
\infty}^{\infty}{{d^9}{p_i}}\biggr]\nonumber\\
&\times\,\delta({\cal E}-{\sum\limits_{i=1}^{n}{{\cal
E}_{i}}})\,{\delta^9}(\sum\limits_{i=1}^{n}{{\vec p}_{i}})\;,
\end{eqnarray}
where ${\cal M}_o$ is the mass of the lightest (extremal) objects
in the gas.
Now for small charge, the degeneracy of states (3.1) actually
belongs to the class of degeneracies ${\rho_p}({\cal M})$ defined
as follows,
\begin{eqnarray}
{\rho_p}({\cal M})\;\equiv\;f({\cal M})\exp{(b{{\cal M}^p})}\;,
\end{eqnarray}
where $f({\cal M})$ is a polynomial in $\cal M$ and $p={{D-
2}\over{D-3}}$($>1$). Recalling the classic results obtained by
Frautschi \cite{Fr71}, the dominant configuration for such a case
is the one for which a single (say the $n^{\rm th}$) black object
carries most of the energy while the n-1 others carry energies
${{\cal E}_i}\,=\,{{\cal M}_o}$ (i=1,...,n-1). So ${{\cal
E}_n}\,=\,{\cal E}-(n-1){{\cal M}_o}$. At high energy $\cal E$,
Eq. (3.4) therefore becomes,
\begin{eqnarray}
{\Omega_n}\;\simeq\;{{\biggl[{V\over{(2\pi)^9}}\biggr]}^n}{1\over{\Gamma(n+1)}}\
M}_o}){{\rho_{B.B.}}({{\cal M}_o})}^{n-1}\;.
\end{eqnarray}

The most probable configuration ${\Omega_N}({\cal E},Q,V)$ is the
one satisfying the following condition,
\begin{eqnarray}
{{\biggl[{{d{\Omega_n}({\cal
E},Q,V)}\over{dn}}\biggr]}_{n=N({\cal E},Q,V)}}\;=\;0\;.
\end{eqnarray}

In complete analogy with the case of the 4 dimensional dilaton
black hole \cite{BL92}, we find the following solution for N,
\begin{eqnarray}
\Psi(N+1)\;=\;\ln{\biggl[{{cV}\over{(2\pi)^9}}\biggr]}\,+\,{S_E}({{\cal
M}_o},Q)
M}_o},Q)\;,
\end{eqnarray}
where $\Psi(z)$ is the psi function and in which use has been
made of the relation (2.32). At high energy (${\cal E}>>(N-
1){{\cal M}_o}$), we get the following approximate relation,
\begin{eqnarray}
\Psi(N+1)\;\sim\;\ln{\biggl[{{cV}\over{(2\pi)^9}}\biggr]}\,-
\,{{4\pi}\over{D-3}}{{\biggl[{{8\Gamma({{D-1}\over2})}\over{(D-
2){\pi^{{D-1}\over2}}}}\biggr]}^{1\over{D-3}}}{{\cal M}_o}{{\cal
E}^{1\over{D-3}}}\,+\,{\cal O}({Q^2})\;.
\end{eqnarray}

It is easy to see that ${{\partial N}\over{\partial{\cal E}}}<0$
and so the most probable configuration at high energy is again
the one for which N is as small as possible, reaching N=1 at a
high energy "ionization point" ${\cal E}_c$. So, in complete
analogy with the case of the 4-dimensional dilaton black hole,
the most probable equilibrium configuration of a gas of black
(10-D)-branes in 9-dimensional space is described as follows,
\begin{eqnarray}
(N-1){{\cal M}_o}<<{\cal E}<<{{\cal
E}_c}\;\;&&(N>>1)\;;\nonumber\\ && \\
{\cal E}\,=\,{{\cal E}_c}\;\;&&(N=1)\;.\nonumber
\end{eqnarray}

The critical energy ${\cal E}_c$ is determined by the formula,
\begin{eqnarray}
{{\cal M}_o}{\beta_H}({{\cal E}_c})\;=\;{S_E}({{\cal
M}_o},Q)\,+\,\ln{\biggl[{{cV}\over{(2\pi)^9}}\biggr]}\,-
\,\Psi(2)\;.
\end{eqnarray}

An approximate solution for small charge is as follows,
\begin{eqnarray}
{{\cal E}_c}\;\simeq\;{{\biggl({{D-3}\over{4\pi{{\cal
M}_o}}}\biggr)}^{D-3}}{{(D-2){\pi^{{D-
3}\over2}}}\over{8\Gamma({{D-
1}\over2})}}{{\biggl[\ln{\biggl({{cV}\over{(2\pi)^9}}\biggr)}\,-
\,\Psi(2)\biggr]}^{D-3}}\;.
\end{eqnarray}

Now since $\Omega({\cal E},Q,V)\sim{\Omega_N}({\cal E},Q,V)$, Eq.
(3.6) shows that the statistical bootstrap condition is trivially
satisfied at ${\cal E}={{\cal E}_c}$ (N=1) since there is a
single object in the gas. We remark that for ${\cal E}>{{\cal
E}_c}$, there is no equilibrium configuration.

The total entropy of the gas is written as follows,
\begin{eqnarray}
S({\cal E},Q,V)\;&\simeq&\;\ln{{\Omega_N}({\cal
E},Q,V)}\;=\;N\ln{\biggl[{{cV}\over{(2\pi)^9}}\biggr]}\,-
\,\ln{\Gamma(N+1)}\nonumber\\
&&+\,{S_E}({\cal E}-(N-1){{\cal M}_o},Q,V)\,+\,(N-1){S_E}({{\cal
M}_o},Q)\;.
\end{eqnarray}

The microcanonical temperature is given by,
\begin{eqnarray}
\beta({\cal E},Q,V)\;&\equiv&\;{{dS({\cal E},Q,V)}\over{d{\cal
E}}}\;=\;{{\partial{S_E}({\cal E}-(N-1){{\cal
M}_o},Q,V)}\over{\partial{\cal E}}}\nonumber\\
&&=\;{\beta_H}({\cal E}-(N-1){{\cal M}_o},Q,V)\;,
\end{eqnarray}
with $N({\cal E},Q,V)$ given by Eq. (3.8) and where Eq. (3.7) has
been used.

Therefore, as in black hole statistical mechanics in 4
dimensions, the microcanonical temperature of a gas of black (10-
D)-branes in 10 dimensions is the same as the Bekenstein-Hawking
temperature of the most massive black object in the gas.

The microcanonical specific heat is now given as,
\begin{eqnarray}
{C_V}\;=\;-\,{\beta^2}{{d{\cal E}}\over{d\beta}}\;.
\end{eqnarray}
Explicit evaluation yields,
\begin{eqnarray}
{C_V}({\cal E},Q,V)\;=\;{{C_V^{\rm (Hawking)}}({\cal E}-(N-
1){{\cal M}_o},Q)}\,{{\biggl[1\,-\,{{\cal M}_o}
{{\partial N}\over{\partial{\cal E}}}\biggr]}^{-1}}\;,
\end{eqnarray}
where,
\begin{eqnarray}
{C_V^{\rm (Hawking)}}\;\equiv\;-\,{\beta_H^2}{{\partial{\cal
E}}\over{\partial{\beta_H}}}\;.
\end{eqnarray}
At high energy, as argued previously, ${{
\partial N}\over{\partial{\cal E}}}<0$ and so the sign of the
microcanonical specific heat is determined by that of the Hawking
specific heat.

The case of neutral (Q=0) black branes is somewhat different as
the extreme limit is massless (${{\cal M}_o}=0$). The degeneracy
of states for such objects reads as follows,
\begin{eqnarray}
\rho_{B.B.}({\cal M})\;\simeq\;c\,\exp{\bigl[\sigma(D){M^{{D-
2}\over{D-3}}}\bigr]}\;.
\end{eqnarray}
At high energy, one finds the following expression for the
corresponding microcanonical density of states,
\begin{eqnarray}
\Omega({\cal
E},V)\;\simeq\;{{\biggl[{{V}\over{(2\pi)^9}}\biggr]}^N}
{{c^{N-1}}\over{\Gamma(N+1)}}\,{{\rho_{B.B.}}({\cal E})}\;,
\end{eqnarray}
which corresponds to a gas consisting of a single supermassive
black object and (N-1) massless others. Therefore we have ${\cal
E}={\cal M}$. Now the most probable configuration is again
determined from the following condition,
\begin{eqnarray}
{{\partial\Omega}\over{\partial N}}\;=\;0\;.
\end{eqnarray}
The solution to this equation is given by Eq. (3.8) with ${{\cal
M}_o}=0$,
\begin{eqnarray}
\Psi(N+1)\;=\;\ln{\biggl[{{cV}\over{(2\pi)^9}}\biggr]}\;.
\end{eqnarray}
Unlike the charged case, we find that the most probable number of
objects in the neutral gas becomes effectively independent of
energy in the high energy domain.

The total entropy of this system is given as follows,
\begin{eqnarray}
S({\cal
E},V)\;\simeq\;N\ln{\biggl[{{cV}\over{(2\pi)^9}}\biggr]}\,-
\,\ln{\Gamma(N+1)}\,+\,{S_E}({\cal E},Q=0)\;.
\end{eqnarray}
The corresponding microcanonical temperature is obtained as
follows,
\begin{eqnarray}
\beta({\cal E},V)\;=\;{\beta_H}({\cal E},Q=0)\;.
\end{eqnarray}
The microcanonical specific heat is negative,
\begin{eqnarray}
{C_V}({\cal E},V)\;=\;-4\pi{{\biggl[{{8\Gamma({{D-
1}\over2})}\over{(D-2){\pi^{{D-3}\over2}}}}\biggr]}^{1\over{D-
3}}}\,{{\cal E}^{{D-2}\over{D-3}}}\;,
\end{eqnarray}
a situation analogous to Schwarzschild black holes in 4
dimensions \cite{BL92}.

As is clear from Eq. (3.19), unlike the case of a gas of charged
black objects, the statistical bootstrap condition can be met for
the neutral gas provided,
\begin{eqnarray}
{{\biggl[{{cV}\over{{2\pi}^9}}\biggr]}^N}{1\over{\Gamma(N+1)}}\;=\;c\;.
\end{eqnarray}
Again, this is due to the fact that extreme neutral black
(10-D)-branes are massless.

Finally, for all cases treated in this section, the
microcanonical equation of state of a gas of black (10-D)-branes
in 10 dimensions is found to be identical to that of an ideal
gas, namely,
\begin{eqnarray}
\beta P\;=\;{N\over V}\;.
\end{eqnarray}
This is consistent with the fact that we neglected collision
processes.

\section{DISCUSSION}
\label{sec:disc}

In this work, we presented a generalization of previous
considerations on black hole statistical mechanics to the case of
black (10-D)-brane solutions recently discovered by Horowitz and
Strominger \cite{HS91}.

The results found here are somewhat similar to those found in the
case of 4-dimensional black holes \cite{BL92}, except that, to
leading order in charge expansion, the Euclidean action
(Bekenstein-Hawking entropy) behaves like $M^{{D-2}\over{D-3}}$
where M is the mass density per unit (10-D)-volume.

As in the case of the 4-dimensional Schwarzschild black holes,
neutral black branes also satisfy the statistical bootstrap
condition, a fact related to the massless nature of the extreme
limit. Also in parallel with Schwarzschild black holes, should we
consider quantum black brane scattering processes, the total
number of open channels, as a simple consideration would show,
actually grows precisely in parallel with the degeneracy of
states as the center of mass energy is increased. Therefore,
\begin{eqnarray}
N(m)\;=\;\sum_{n=1}^{\infty}\,{N_n}(m)\;\sim\;{\rho_{B.B.}}(m)\;;\;(m\rightarrow
\end{eqnarray}
Again, it is then plausible to argue that neutral black brane
scattering amplitudes satisfy the duality (crossing) symmetry
characteristic of string theories. Like strings, they perhaps
belong to a class of conformal field theories, e.g. N-branes.

Calculations of the degeneracy of states of higher dimensional
structures such as quantum N-branes have been presented almost
two decades ago by a few authors \cite{FHJ73,DNT74,SV75} and
happily re-discovered more recently by the authors of
Ref.~\cite{Al91}. According to these calculations, the asymptotic
behavior of the degeneracy of states of N-branes at large energy
(mass) is given as follows,
\begin{eqnarray}
{\rho_{N}}({\cal E})\;\propto\;\exp{[b\,{{\cal
E}^{{2N}\over{N+1}}}]}\;,
\end{eqnarray}
where N is the dimensionality of the extended objects. The above
asymptotic behavior seems to be valid in any space-time dimension
although the parameter b may be dependent upon the space-time
dimensionality. It is interesting to compare this result with our
Eq. (3.18) describing the behavior of the degeneracy of states of
a quantum neutral black (10-D)-brane in 10 space-time dimensions
(or analogously a neutral black hole in D dimensions). One finds
the following relation,
\begin{eqnarray}
N\;=\;{{D-2}\over{D-4}}\;.
\end{eqnarray}
Only three solutions for integer N exist in the allowed range
$4\leq D\leq 10$, namely $N=2\,(D=6)$, $N=3\,(D=5)$ and
$N=\infty\,(D=4)$. We then arrive at the conclusion that black
4-branes in 10 dimensions are 2-branes, black 5-branes are
3-branes and black 6-branes are $\infty-{\rm branes}$, or
analogously 6-dimensional Schwarzschild black holes are 2-branes,
the 5-dimensional ones are 3-branes and finally 4-dimensional
Schwarzschild black holes are $\infty-{\rm branes}$. This last
result, already pointed out in Ref.~\cite{Al91}, may
somewhat come as a surprise in view of the usual membrane
(2-brane) viewpoint on 4-dimensional black holes \cite{PT86}.

The considerations presented here with regard to the
interpretation of the semiclassical approximation of the
Euclidean path integral are completely quantum mechanical.
Periodicity of the instanton solutions certainly does not
constrain one to a thermal interpretation. Actually, such an
interpretation would lead to the following paradoxical situation,
namely, at least for neutral black objects (including black
holes), should the semiclassical approximation of the path
integral be interpreted as the canonical partition function of a
gas of black objects at inverse temperature $\beta_H$, then its
corresponding (statistical mechanical) density of states would be
the same as the one obtained from a quantum system with
degeneracy of states $\rho_{B.B.}$ obeying the bootstrap
condition. However, we know that the thermal partition function
for a gas of objects with such a degeneracy of states does not
exist for any finite temperature.

Clearly, the resolution of this paradox lies in the fact that the
gas does not achieve thermal equilibrium (the microcanonical
specific heat is negative) and consequently the microcanonical
and canonical ensembles are not equivalent. It is well known
that in this situation, the saddle point approximation fails when
passing from one ensemble to the other (recall that the canonical
partition function is the Laplace transform of the density of
states). For such systems of course, one should trust the
microcanonical ensemble because it is more fundamental in
ensemble theory.

Although formally infinite for all temperatures, it is however
possible to extract information from the canonical partition
function by evaluating it in the convergence domain of certain
parameters. Once the integration (over mass) has been performed
these parameters are then analytically continued back to their
original values. This procedure usually produces a finite but
complex partition function. The nucleation rate (decay rate per
unit volume) of the black objects gaseous phase is then simply
related to the imaginary part of the corresponding free energy
\cite{La6769}, a calculation not unlike the determination of the
decay rate of the false vacuum when dealing with a complex
effective potential in ordinary quantum field theory [27,28].
Analogous situations also occur in string theories
\cite{Mar88,AT88,YL90}.

\acknowledgments

We are grateful to G. Venturi for making us aware of the work of
Refs.~\cite{FHJ73,DNT74,SV75} as well as P. Townsend for guidance
in the litterature on p-brane theories.

This research was supported in part by the U.S. Department of
Energy under Grant No. DE-FG05-84ER40141 and in part by the Texas
National Research Laboratory Commission under Grant No. RCFY92-
117.


\begin{references}
\bibitem{tH90} G. 't Hooft, Nucl. Phys. {\bf B335}, 138 (1990) ;
and references therein.
\bibitem{Dt85} T. Dray and G. 't Hooft, Nucl. Phys. {\bf B253},
173 (1985).
\bibitem{Ve91} H. Verlinde and E. Verlinde, PUPT-1279 (1991).
\bibitem{Co92} P. H. Cox, B. Harms and Y. Leblanc, UAHEP-9213
(1992).
\bibitem{Di88} B. L. Dingus {\it et al.}, Phys. Rev. Lett. {\bf
61}, 1906 (1988).
\bibitem{Be73} J. D. Bekenstein, Phys. Rev. {\bf D7}, 2333
(1973).
\bibitem{Ha75} S. W. Hawking, Comm. Math. Phys. {\bf 43}, 199
(1975) ; Phys. Rev. {\bf D13}, 191 (1976) ; and references
therein.
\bibitem{CPW91} S. Coleman, J. Preskill and F. Wilczek, Mod.
Phys. Lett. {\bf A6}, 1631 (1991).
\bibitem{Pr91} J. Preskill, P. Schwarz, A. Shapere, S. Trivedi
and F. Wilczek, IASSNS-HEP-91/34 (1991).
\bibitem{SC91} S. Coleman, J. Preskill and F. Wilczek,
IASSNS-HEP-91/64 (1991) ; and references therein.
\bibitem{Ho91} C. F. E. Holzhey and F. Wilczek, IASSNS-HEP-91/71
(1991).
\bibitem{BL92} B. Harms and Y. Leblanc, to appear in Phys. Rev.
{\bf D46} (1992).
\bibitem{GPY81} D. J. Gross, R. D. Pisarski and L. G. Yaffe, Rev.
Mod. Phys. {\bf 53}, 43 (1981).
\bibitem{GPY82} D. J. Gross, M. J. Perry and L. G. Yaffe, Phys.
Rev. {\bf D25}, 330 (1982).
\bibitem{Hg70} R. Hagedorn, Nuovo Cim. Suppl. {\bf 3}, 147
(1970).
\bibitem{Fr71} S. Frautschi, Phys. Rev. {\bf D3}, 2821 (1971).
\bibitem{Ca72} R. D. Carlitz, Phys. Rev. {\bf D5}, 3231 (1972).
\bibitem{HS91} G. T. Horowitz and A. Strominger, Nucl. Phys. {\bf
B360}, 197 (1991).
\bibitem{GM88} G. W. Gibbons and K. Maeda, Nucl. Phys. {\bf
B298}, 741 (1988).
\bibitem{La6769} J. S. Langer, Ann. Phys. (N.Y.) {\bf 41}, 108
(1967) ; {\bf 54}, 258 (1969).
\bibitem{SeeSC} e.g. see Ref.~\cite{SC91}.
\bibitem{FHJ73} S. Fubini, A. J. Hanson and R. Jackiw, Phys. Rev.
{\bf D7}, 1732 (1973).
\bibitem{DNT74} J. Dethlefsen, H. B. Nielsen and H. C. Tze, Phys.
Lett. {\bf B48}, 48 (1974).
\bibitem{SV75} A. Strumia and G. Venturi, Lett. Nuovo Cim. {\bf
13}, 337 (1975).
\bibitem{Al91} E. Alvarez and T. Ortin, CERN-TH.6345/91 (1991).
\bibitem{PT86} R. H. Price and K. S. Thorne, Phys. Rev. {\bf
D33}, 915 (1986).
\bibitem{Col 77} S. Coleman, Phys. Rev. {\bf D15}, 2929 (1977) ;
{\bf D16}, 1248 (1977).
\bibitem{WW87} E. J. Weinberg and A. Wu, Phys. Rev. {\bf D36},
2474 (1987).
\bibitem{Mar88} N. Marcus, University of Washington Report No.
40423-21 (1988).
\bibitem{AT88} K. Amano and A. Tsuchiya, TIT/HEP-132 (1988).
\bibitem{YL90} Y. Leblanc, M. Knecht and J. C. Wallet, Phys.
Lett. {\bf B237}, 357 (1990).
\end{references}
\end{document}